\documentclass[final]{elsarticle}

\usepackage[english]{babel}
\usepackage[%
colorlinks=true,
linkcolor=blue,
citecolor=blue,
urlcolor=blue
]{hyperref}

\usepackage{mathtools}
\usepackage{amsmath}
\usepackage{amsthm}
\usepackage{amsfonts}
\usepackage{float}
\usepackage{graphicx}
\usepackage{caption}
\usepackage{subcaption}
\usepackage{physics}
\usepackage{url}
\usepackage{epstopdf}
\usepackage{lipsum}
\usepackage[title]{appendix}
\usepackage{titlesec}
\usepackage{gensymb}
\usepackage[separate-uncertainty = true]{siunitx}
\usepackage{colortbl}
\usepackage{wrapfig}
\usepackage{setspace}
\usepackage{amssymb}
\usepackage[ampersand]{easylist}
\usepackage{array}
\usepackage{cancel}
\usepackage[T1]{fontenc}
\usepackage[utf8]{inputenc}
\usepackage[flushleft]{threeparttable}
\usepackage{soul}
\usepackage{xcolor}
\sethlcolor{yellow}
\usepackage[ margin=1in]{geometry}
\usepackage{lmodern}

\graphicspath{ {./Images/}}

\journal{journal for review}

\begin{document}

\title{Mesoscale shock structure in particulate composites}

\author[add1]{Suraj Ravindran\corref{mycorrespondingauthor}}
\cortext[mycorrespondingauthor]{Corresponding author}
\ead{sravi@umn.edu}
\author[add2]{Vatsa Gandhi}
\author[add2]{Barry Lawlor}
\author[add2,add3]{Guruswami Ravichandran}

\address[add1]{Department of Aerospace Engineering and Mechanics, University of Minnesota, Minneapolis, MN 55455, USA}
\address[add2]{Division of Engineering and Applied Science, California Institute of Technology, Pasadena, CA 91125, USA}
\address[add3]{Jio Institute, Ulwe, Navi Mumbai, Maharashtra 410206, India}

\begin{abstract}
Multiscale experiments in heterogeneous materials and the knowledge of their physics  under shock compression are limited. This study examines the multiscale shock response of particulate composites comprised of soda-lime glass particles in a PMMA matrix using full-field high-speed digital image correlation (DIC) for the first time. Normal plate impact experiments, and complementary numerical simulations, are conducted at stresses ranging from $1.1-3.1$ GPa to eludicate the mesoscale mechanisms responsible for the distinct shock structure observed in particulate composites. The particle velocity from the macroscopic measurement at continuum scale shows a relatively smooth velocity profile, with shock thickness decreasing with an increase in shock stress, and the composite exhibits strain rate scaling as the second power of the shock stress. In contrast, the mesoscopic response was highly heterogeneous, which led to a rough shock front and the formation of a train of weak shocks traveling at different velocities. Additionally, the normal shock was seen to diffuse the momentum in the transverse direction, affecting the shock rise and the rounding-off observed at the continuum scale measurements. The numerical simulations indicate that the reflections at the interfaces, wave scattering, and interference of these reflected waves are the primary mechanisms for the observed rough shock fronts.   
\end{abstract}

\maketitle

\section{Introduction} \label{introduction}

Heterogeneous materials, such as composites, have widespread use in structural, propulsion, and armor applications \cite{forquin2017brittle, baer2002modeling,Jannotti_2021} due to their ability to sustain intense shock waves. Specifically, particulate-reinforced composites are prevalent in developing energetic materials, such as polymer-bonded explosives (PBX) \cite{baer2002modeling}, materials for protective structures (e.g. potting compounds used to provide shock absorption for electronic devices) \cite{MitchellPotting1981}, and infrastructural materials, such as ultra-high performance concrete \cite{neel2018compaction}, which all necessitate the dissipation of shock to mitigate any damage or prevent any accidental detonations. 

One of the main advantages of these composite materials in dynamic impact applications is their ability to disperse the shock. Shock waves are a discontinuity resulting in an abrupt, and irreversible change in field variables such as particle velocity, stress, and strain. While the shock structure is often assumed to be infinitely thin, in reality, it has a finite width which is said to be controlled by viscosity. For homogeneous materials, Swegle and Grady \cite{grady1981strain,swegle1985shock} have shown the relation between stress ($\sigma$) and strain rate  across a steady shock front is assumed to follow the relation, $\dot{\varepsilon} \propto \sigma^{4}$. Experimental investigations have consistently shown agreement with this relation for homogeneous materials \cite{swegle1985shock,crowhurst2011invariance}, known as the Swegle-Grady relation. Additionally, Molinari and Ravichandran \cite{molinari2004fundamental} have developed general analytical solutions for the steady shock structure in homogeneous metals (e.g. aluminum) and found the Swegle-Grady relation to hold but with a slightly different strain rate sensitivity. Regardless, the physical basis for why the strain rate scales to the fourth power of the stress is still not fully understood. While homogeneous materials follow this particular relation, the shock structure of heterogeneous materials deviates from this result. Some have found the relation to hold for alumina filled epoxy (ALOX)\cite{setchell2005shock} and tungsten carbide filled epoxy (WCE) \cite{vogler2010dynamic}, while others report power laws with exponents less than four (e.g. $\dot{\varepsilon} \propto \sigma^{2-3}$ for particulate composites \cite{rauls2020structure,vogler2012scaling}, $\dot{\varepsilon} \propto \sigma^{2.4}$ for layered composites \cite{vogler2012scaling}, and $\dot{\varepsilon} \propto \sigma^{1}$ for granular media \cite{vogler2012scaling,vogler2007ceramic,brown2007sand}). This is a clear indication that alternate mechanisms play a role in the shock front propagation in heterogeneous materials which make them favorable for applications requiring wave dissipation.

The fundamental shock structure of layered composites, granular media, and particulate composites is shown to largely be governed by a particular characteristic length. For layered composites, work by Molinari and Ravichandran \cite{molinari2006modeling} and Zhuang \textit{et al.} \cite{zhuang2003experimental} demonstrated shock structure is primarily governed by the cell size of a laminate, and viscosity plays a minor role. In the case of granular media, it has been shown that the grain diameter governs the compaction wave thickness \cite{borg2008mesoscale} while, the shock thickness of a particulate composite is expected to be on the order of the particle size \cite{rauls2020structure}. The primary observation regarding the shock structure in composites is governed by varying wave scattering mechanisms. Specifically, for particulate composites, many theories suggest that wave reflection at the matrix-particle interface is the primary mechanism for shock structure elongation \cite{rauls2020structure,grady1998scattering,chen2004analytical}, while others have proposed material flow \cite{bober2019situ}, particle-particle contact \cite{hurley2018situ}, and particle deformation due to fluid-particle interaction \cite{bober2022observations} as significant mechanisms.

These scattering mechanisms in particulate composites result in complex wave interactions influencing the propagating shock wave and thus, understanding the dynamic shock structure of these heterogeneous materials requires characterizing not only the macroscopic response but also resolving the meso-scale features. Significant fundamental research has been performed to systematically study the influence of parameters such as impedance mismatch ratio between matrix and particle \cite{rauls2020structure}, size distribution of particles \cite{rauls2020structure,vogler2012scaling}, composition percentages, particle order, particle geometry \cite{baer2002modeling,setchell2005shock}, and shock pressure on the wave dispersion mechanisms. From a numerical standpoint, Vogler \textit{et al.} \cite{vogler2012scaling} performed meso-scale simulations on many common particulate composites, and showed that increasing particle size decreased the strain rate in the material for a given stress. They also suggest that ordered particulate composites exhibit a stiffer response with lower velocity rise times. Similarly, Baer \cite{baer2002modeling} has demonstrated the ordered structure is shown to generate more shock wave resonance (i.e. larger amplitude of stress fluctuations) when compared with random, disordered particle distribution. The author also compared the response of spherical particles and cubic particles, finding that spherical particles generate more resonance due to the non-planar matrix-particle interfaces. Bober \textit{et al.} \cite{bober2020description} investigated a polymer matrix filled with tungsten particles, and found it to exhibit a shock structure analogous to that of a homogeneous viscoelastic material, comprised of an initial rapid acceleration before a slow rounding off as the velocity approaches an equilibrium value. They define a time constant representing the rise time of initial acceleration, and showed that increasing tungsten content produced a smaller time constant, while increasing impact velocity initially produced a larger time constant up until a threshold of impact velocity, after which the time constant decreases with rising impact velocity.

A number of experimental efforts have been able to provide qualitative observations via imaging techniques such as x-ray phase contrast imaging \cite{wagner2020highspeed} and radiography \cite{bober2022observations}. Ultrafast radiographs of metallic particles suspended in epoxy show to time evolution of particle deformation as a function of particle strength \cite{bober2022observations}. Numerical models are utilized to match the morphology, and fitting across various impact velocities enables the model to predict the transition from strength-related regime to a hydrodynamic regime. The work also elucidates the importance of shock-particle interaction, stagnation pressure, and vorticity on the deformation of the particle. Using x-ray phase contrast imaging, a new DIC technique is developed to measure internal strains in an additively manufactured energetic material simulant \cite{wagner2020highspeed}. Though the strain response reported is essentially isotropic, it is proposed that the method could be used to study the spatial variation of the shock response in heterogeneous media.

Quantitatively, laser interferometry \cite{barker1965interferometer,barker1972laser} has been the favored velocity measurement technique for plate impact experiments. Through a combination of experiments and simulations, many studies probe the macroscale properties of particulate composites such as tungsten carbide filled epoxy \cite{vogler2010dynamic}, aluminum-iron oxide-epoxy \cite{jordan2007equation}, and multi-constituent epoxy-metal or -ceramic particulate composites \cite{jordan2011shock,jordan2010shock}. Specifically, Rauls and Ravichandran \cite{rauls2020structure} developed a model composite material to study shock disruption and scattering mechanisms using multiple PDV measurements to obtain higher spatial resolution and mitigate errors associated with the spatially heterogeneous response. They determine that increasing heterogeneity (i.e. increasing bead size) increases the shock width, but bead sizes on the order of the shock thickness of the pure matrix material are much less effective shock disruptors when compared with larger bead sizes. Additionally, they observe primary and secondary shock reflections at the matrix-particle interfaces govern the shock width, and bead size and location influence the behavior of these reflected waves. Finally, the authors developed an empirical relation showing the shock thickness scales linearly with bead size, with a slope close to 1. The multiple-PDV measurement approach improves measurement fidelity when establishing important shock structure relationships, and provides new insights into the heterogeneous shock response of particulate composites, motivating the need for further improved spatial resolution.

As stated earlier, because the shock response of these composite materials is inherently heterogeneous, point measurements are unable to fully represent the average behavior or statistical distribution of local responses. To this end, many aforementioned studies have reported the averaged response over multiple points, and other interferometry techniques such as line-ORVIS \cite{bloomquist1983optically,baumung1996orvis,trott2000measurements} would further extend the spatial resolution beyond a few point measurements. However, even a line of measurements would be unable to fully capture the three-dimensional, heterogeneous, statistical nature of the shock response. Unfortunately, these experimental limitations have made it difficult to fulfill the obvious need for meso-scale measurements. In response, many have depended on point-wise measurements in conjunction with simulations to draw meaningful conclusions. 

The improvements in high speed imaging and full-field quantitative visualization techniques, such as digital image correlation (DIC), enable \textit{in-situ} meso-scale experimental measurements. Ravindran, Gandhi, \textit{et al.} \cite{DICPaper} have recently developed an extension of high speed stereo-DIC to carryout full-field free surface velocity measurements during plate impact experiments. The main objective of this study is to (1) resolve the complete meso-scale heterogeneous response of these model particulate composites \cite{rauls2020structure} using full-field velocity measurements and (2) explore the fundamental mechanisms regarding the strain rate scaling at varying stresses. The DIC technique is utilized to perform plate impact experiments on a model particulate composite of glass beads embedded in PMMA, where the spatial resolution is on the order of the bead size--sufficient to resolve the heterogeneity caused by the beads--and the temporal resolution is sufficient to resolve the rise time of the shock response. The technique will be described in section \ref{methods}, along with details on normal plate impact experiments on particulate composites. Next, full-field results will be presented in section \ref{results} in light of scattering, out of plane oscillations, and in-plane heterogeneity. Traditional metrics such as the stress-strain rate power law will be discussed, in addition to new abilities to measure the statistical aspects of the shock response.

\section{Materials and Methods} \label{methods}

\subsection{Materials} \label{materials}
Plate impact experiments coupled with ultra-high speed digital image correlation (DIC) were conducted using a powder gun with a 3 meter long and 38.7 mm bore diameter slotted barrel. The targets consisted of particulate composites composed of PMMA matrix embedded with 1 mm diameter soda-lime glass (SLG) beads (23\% volume fraction), and were glued circumferentially to an aluminum ring used for measuring impact tilt and triggering data capture. The impactor materials varied between copper (Cu), magnesium AZ31B (Mg), and aluminum 7075 (Al) depending on the desired shock stresses. The material properties, obtained from literature, and the sample geometry are summarized in Tables \ref{tab:properties} and \ref{tab:Overview} respectively. More details are explained in the subsequent sections.

\subsection{Sample Preparation} \label{sampleprep}

The glass composite sample consisting of PMMA matrix with SLG particles was prepared as follows. First, an acrylic monomer powder (Metlab Corporation, Niagara Falls, New York) was mixed with the soda-lime glass beads (Cospheric Corporation, Goleta, California) of 1 mm. The curing agent was added next and the mixture was stirred until it became viscous to prevent any settling of the glass beads. Lastly, this mixture was poured into a 30 mm diameter aluminum mold with a 3 MPa applied pressure to compact the mixture. The sample was cured in the mold at room temperature for 1 hour after which it was removed. 

\begin{figure}[ht]
	\centering
	\includegraphics[width=1\textwidth]{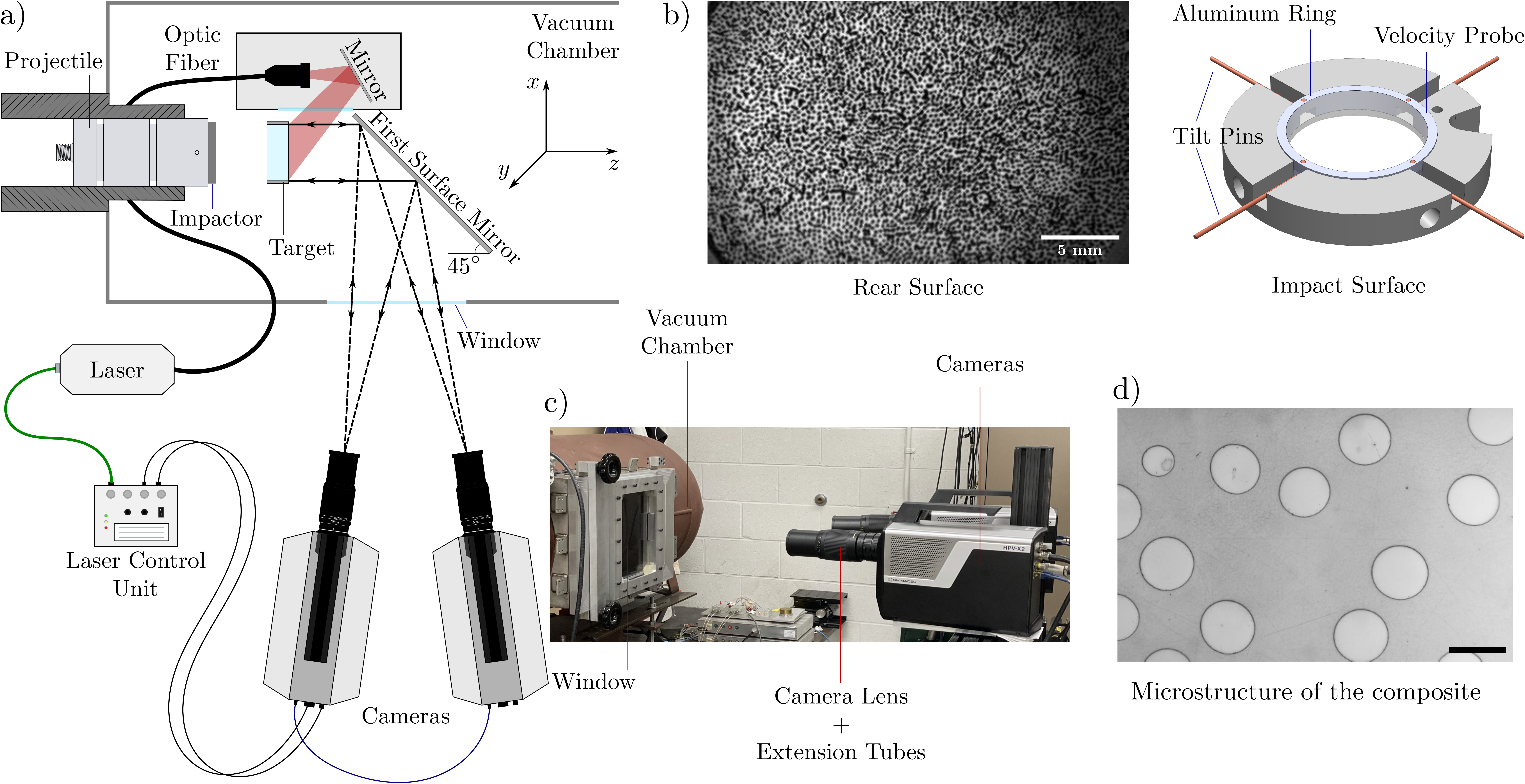}
	\caption{a) Schematic of the experimental setup for full-field free surface velocity characterization of particulate composites using digital image correlation (DIC), b) overview of the target and a representative speckle pattern on the free surface of the sample, c) the high-speed camera arrangement for stereo DIC set up for imaging through the transparent window of the chamber, d) image of the microstructure of a representative composite sample. The scale bar is $1$ mm.}
	\label{fig:Exp}
\end{figure}

To ensure plane wave propagation upon impact, plate impact experiments require precise sample preparation. Specifically, both the sample and the flyer must be flat and aligned parallel to one another. Hence, once the composite has been cured, the sample is lapped flat on both sides while exercising caution to avoid any delamination of the SLG beads from the matrix. Any voids formed through delamination of the glass particles could potentially influence the experimental results. To accommodate this issue, the lapping is kept to a minimum and consequently, the tolerances are relaxed slightly such that a parallelism of 20 $\mu$m on the glass composite is acceptable. The flatness variation of the sample is inspected by visualizing the circular Fizeau fringes using 550 nm monochromatic light and an optical flat. A variation on the order of 0.5 $\mu$m was deemed acceptable. The specimen tolerances on the flyer, however, are maintained at standard values for typical plate impact experiments such that it is lapped flat to within 0.5 $\mu$m variation across the plate surface and parallel to within 10 $\mu$m. Once the plates satisfy the required criteria, the flyer is glued onto the projectile and an aluminum ring is glued concentrically to the glass composite to facilitate triggering diagnostics. A series of four electrical shorting pins are glued into equally spaced slots on the aluminum ring as shown in Figure \ref{fig:Exp}. The target is then lapped flat again on the impact surface to ensure the glass composite, aluminum ring, and the copper shorting pins are all flush together. The shorting pins are connected to a digital circuit that is linked to an oscilloscope. Upon impact, the imaging diagnostics are triggered when at least two pins make contact with the impactor, while the contact times of the shorting pins are used to determine the planarity of the impact. After the target has been fully assembled, a thin coat of white paint is applied to the rear surface of the target and the paint is left to dry for 24 hours before an isotropic speckle pattern is applied to the sample. The speckle is applied using a fine tip (30 $\mu$m) pen after which it is left to dry for 24 hours to prevent any smearing during the experiment. Once the speckles are fully dry, the sample is glued onto a target holder, affixed on a six-axis gimbal, and placed in the chamber along with the projectile. The six-axis gimbal allows rotation and translation of the target in all directions and thus is critical for alignment which is discussed in the next section.

\subsection{Experimental Setup} \label{plateimpact}

In a plate impact experiment, a projectile is accelerated along a barrel and the impact between the flyer and target generates a plane shock wave propagating in both materials. In addition to the shock waves generated at the impact interface, the traction free boundary conditions at the edges of the samples result in a boundary wave propagating radially towards the center of the sample at the longitudinal wave speed. Generally, the diagnostics are conducted before the arrival of the boundary waves to ensure the material is under a uniaxial strain state. However, due to the introduction of SLG inclusions in PMMA, the material may not necessarily be under a state of uniaxial strain. Nonetheless, the results are analyzed within this time window prior to the boundary wave arrival at the imaging field of view on the rear surface. 

To generate a plane wave structure, the first step was to ensure the sample and flyer are lapped flat and parallel, and the second step deals with the final alignment before an experiment within the chamber where impact occurs. For the inital alignment, the gimbal was translated in the horizontal and vertical directions to concentrically align the impactor and target. Next, using a pair of optically flat mirrors attached to the surface of the sample and impactor and an auto-collimator, the two interfaces are angled with respect to each other to within 0.2 milliradian (mrad) tilt \cite{kumar1977optical}. It is important to note the projectile has a series of keys to prevent any rotation as it travels along the slotted barrel and thus the alignment remains undisturbed during an experiment. 

Photonic doppler velocimetry (PDV) is used to measure impact velocity. However, unlike conventional plate impact experiments which use point-based laser interferometry for out-of-plane free surface velocity measurements, here we conduct full field DIC analysis \cite{DICPaper}. The images of the speckles on the rear free-surface of the target are captured using two Shimadzu HPV-X2 cameras (Shimadzu, Kyoto, Japan) with a stereo angle ranging from $11^{\circ}-16^{\circ}$ and at a framing rate of 10,000,000 fps. Two 300 mm lenses with a 48 mm extension tube were used to maximize the target free surface in the imaging frames and ensure 4-6 pixels per speckle. This resulted in a 68-74 $\mu$m/pixel resolution. Since the chamber is under vacuum during an experiment, the cameras look through a $12.5$ mm thick polycarbonate window. In order to minimize error, the two cameras were calibrated through this window using a $9\times9$ calibration grid with a 2 mm grid spacing. Additionally, the rear surface was visualized through a $50\times 100$ mm first-surface mirror which was placed in the chamber behind the target and angled at $45^{\circ}$ with respect to the loading direction. The target was illuminated using an incoherent Cavilux high-speed laser (Cavitar, Tampere, Finland) with an exposure time of 50 ns synced with the camera shutter. The low exposure times of the laser ensures the capture of still frames during the experiment and thus minimizes motion blurring from high transients produced during shock loading. Due to the low exposure times and camera magnification, it is difficult to sufficiently illuminate the sample from outside the chamber. Thus, the laser lighting was brought inside and placed as close to the sample as possible.

\subsection{Data Processing} \label{processing}
Both the calibration and the deformation images captured during the experiment were processed using the Vic-3D software from Correlated Solutions ( Columbia, South Carolina). The calibration images provide the necessary information to determine the orientations of the two cameras with respect to each other. Regarding the deformation, the DIC analysis is used to extract the three components of the full-field displacement of the sample both in- and out-of-plane from which the corresponding velocities are determined via differentiation. A summary of the experimental resolution and the parameters, such as step and subset size, used for the analysis are shown in Table \ref{tab:DIC}. As shown previously \cite{DICPaper}, the uncertainty in the out-of-plane velocity was approximately 64 m/s while the in-plane velocity uncertainty was much lower, on the order of 8 m/s. To mitigate these uncertainties, a smoothing spline had been fit to the displacement vs. time data which was then differentiated numerically to obtain the velocities. Through this smoothing, the uncertainties were reduced to 45 m/s for the out-of-plane velocity. Additionally, a step size of 1 pixel (px) was used to fully capture the spatial resolution due to the wave scattering and interactions from the glass bead impurities.  

\begin{table}[h]
 	\centering
 	\begin{threeparttable}
 		\setlength{\tabcolsep}{7.5pt}
 		\caption{Experimental resolution and data processing parameters for DIC}
 		\centering
 		\footnotesize
 		\begin{tabular}{cccccc} 
 			\hline \hline
 			\textbf{Experiment} & \begin{tabular}{@{}c@{}}\textbf{Stereo Angle} \\ \textbf{{[}deg{]}}\end{tabular} & \begin{tabular}{@{}c@{}}\textbf{Spatial}\\ \textbf{Resolution} \\ \textbf{{[}$\mu$m/px{]}}\end{tabular}& \begin{tabular}{@{}c@{}}\textbf{Subset Size} \\ \textbf{{[}px $\times$ px{]}}\end{tabular} &  \begin{tabular}{@{}c@{}}\textbf{Step Size} \\ \textbf{{[}px{]}}\end{tabular} & \begin{tabular}{@{}c@{}}\textbf{Temporal}\\ \textbf{Resolution} \\ \textbf{{[}ns{]}}\end{tabular} \\ \hline
 			Cu-GPC           & 15.6       & 68             & $21 \times 21$        & 1    & 100        \\
 			Mg-GPC-1             & 16.1       & 72             & $17 \times 17$        & 1    & 100        \\
 			Mg-GPC-2          & 18.1       & 68             & $21 \times 21$       & 1    & 100        \\
 			Al-GPC             & 15.2       & 68             & $21 \times 21$        & 1  & 100 \\ \hline \hline            
 		\end{tabular}
 		\label{tab:DIC}
 	\end{threeparttable}
 \end{table}
  
\subsection{Numerical Simulations}
Two dimensional finite element simulations were conducted using ABAQUS/Explicit \cite{Abaqus_manual} (Dassault Systemes, Providence, RI) to both design and validate the experiments. A sample with 23\% volume fraction of randomly distributed 1 mm diameter glass beads, similar to the experiment, was replicated in the simulations. These simulations were used to validate the observed macroscopic free surface velocities and the calculated stresses in the experiment. Additionally, they were used to explore the mesoscopic mechanisms observed experimentally to better understand the heterogeneous shock structure. For all simulations, a mesh size of $20$ $\mu$m was used to fully capture the spatial features arising from the shock wave scattering and alternative mechanisms. This mesh size was found by conducting convergence studies by varying the mesh size. 

A plasticity model was implemented to capture the strength behavior of these materials under high normal stresses imposed by the traversing shock wave. The strength of PMMA, aluminum 7075, and AZ31B were modeled using a strain rate dependent Johnson-Cook model while copper was modeled using power law hardening with Cowper-Symmonds strain rate hardening. The SLG beads were treated as elastic-perfectly plastic with a Cowper-Symmonds strain rate hardening model. The Johnson-Cook model and the power law model with Cowper-Symmonds strain rate hardening are shown below, in Equations (\ref{eq:JC}) and (\ref{eq:PL}) respectively. A summary of the strength parameters are shown in Table \ref{tab:Strength}.

\begin{align}
    Y&=\left(A+B \varepsilon_{p}^{n}\right)\left(1+C \ln \frac{\dot{\varepsilon}}{\dot{\varepsilon}_{r e f}}\right)\left(1-\left(\frac{T-T_{R}}{T_{m}-T_{R}}\right)^{m}\right) \label{eq:JC} \\\nonumber\\
    Y&=\left(A+B \varepsilon_{p}^{n}\right)\left(1+\left(\frac{\dot{\varepsilon}_{p}}{D \dot{\varepsilon}_{r e f}}\right)^{p}\right)\left(1-\left(\frac{T-T_{R}}{T_{m}-T_{R}}\right)^{m}\right) \label{eq:PL}
\end{align}

\noindent here, $Y$ corresponds to the yield function of the material, $A$ is the initial yield stress, $B$ is the strain hardening coefficient, and $n$ is the strain hardening exponent. For Johnson-Cook, $C$ corresponds to the strain rate hardening coefficient while for Cowper-Symmonds, $D$ and $p$ correspond to the strain rate hardening coefficient and exponent, respectively. Regarding the temperature term, $m$ corresponds to the temperature softening exponent, $T_R = 300$ K is the reference temperature, and $T_m$ corresponds to the melting temperature for the respective material. However, the temperature term in the yield function were ignored for the simulations. To model the volumetric response of the materials, Mie-Gr\"{u}neisen equation of state was used and the parameters for the linear equation of state, $U_s = C_0 + Su_p$ are displayed in Table \ref{tab:properties}, where $U_s$ and $u_p$ are the shock velocity and particle velocity, respectively, $C_0$ is the bulk sound wave speed, and $S$ is the linear slope parameter. For the SLG beads, a tabular equation of state was used instead, from Joshi et al. \cite{Joshi2021}. 

\begin{table}[t]
    \setlength{\tabcolsep}{7.5pt}
    \captionsetup{justification=centering}
 	\caption{Parameters of the strength model used in the simulations.}
 	\centering
 	\footnotesize
    \begin{tabular}{cccccccccc}
    \hline\hline
    \multicolumn{1}{l}{} & \multicolumn{1}{l}{}                  & \multicolumn{8}{c}{Model parameters (Eqs. (\ref{eq:JC}) and (\ref{eq:PL}))}                                                                                                                         \\ \cline{3-10} 
    Material    & \begin{tabular}{@{}c@{}} Shear Modulus \\ $G$ {[}GPa{]}\end{tabular} & \begin{tabular}{@{}c@{}} $A$ \\ {[}MPa{]}\end{tabular} & \begin{tabular}{@{}c@{}} $B$ \\ {[}MPa{]}\end{tabular} & $n$ & $C$ & $m$ & $p$ & \begin{tabular}{@{}c@{}} $D$ \\ ($\times10^3$)\end{tabular} & \begin{tabular}{@{}c@{}} $\dot{\varepsilon}_{ref}$ \\ {[}1/s{]}\end{tabular} \\ \hline
    PMMA \cite{rai_mechanics_2020}                   & 2.2                                  & 357                    & 90                     & 0.40         & 0.077        & 0.94        & --            & --                           & 1                                  \\
    SLG \cite{Joshi2021}                   & 30.5                                   & 6100                   & --                   & --         & --            & --            & 2            & 500                        & --                                  \\
    OFHC Copper \cite{Ravindran_2021}               & 47.7                                  & 250                    & 100                    & 0.24         & --            & --            & 2            & 500                         & 1                                  \\
    Aluminum 7075 \cite{Brar_2009}                 & 26.9                                   & 546                    & 678                     & 0.71         & 0.024       & 1.56         & --            & --                           & 1        \\
    AZ31B \cite{Sun_2022}                 & 17                                   & 225                    & 168                     & 0.242         & 0.013       & 1.55         & --            & --                           & 0.001        \\ \hline\hline                         
    \end{tabular}
    \label{tab:Strength}
\end{table}

 \begin{table}[h]
 	\centering
 	\begin{threeparttable}
 		\setlength{\tabcolsep}{7.5pt}
 		\caption{Material properties and Mie-Gr\"{u}neisen equation of state parameters}
 		\centering
 		\footnotesize
	\begin{tabular}{cccc}
		\hline \hline
		Material & \begin{tabular}{@{}c@{}} Density, $\rho$ \\ {[}kg/m$^3${]}\end{tabular}  & \begin{tabular}{@{}c@{}}$C_0$ \\ {[}m/s{]}\end{tabular} & $S$ \\ \hline
		PMMA \cite{rauls2020structure}   & 1186                      & 2600                                     & 1.5                                     \\
		SLG \cite{Joshi2021}        & 2480                       & --                                     & --          \\ 
		Cu \cite{Ravindran_2021}   & $8950$                     & 3933                                     & 1.52                                     \\
		Mg \cite{Sun_2022}   & 1770                     & 4520                                     & 1.242                                     \\
		Al \cite{marsh_1980}   & 2804                      & 5200                                     & 1.36                                    \\\hline\hline                          
	\end{tabular}
	\label{tab:properties}
 		\begin{tablenotes}
 			\footnotesize
 			\item **A tabular EOS was used for SLG from \cite{Joshi2021}
 		\end{tablenotes}
 	\end{threeparttable}
 \end{table}

\section{Results and Discussion} \label{results}
\subsection{Macroscale response of particulate composite}
\begin{figure}[ht]
	\centering
	\includegraphics[width=1\textwidth]{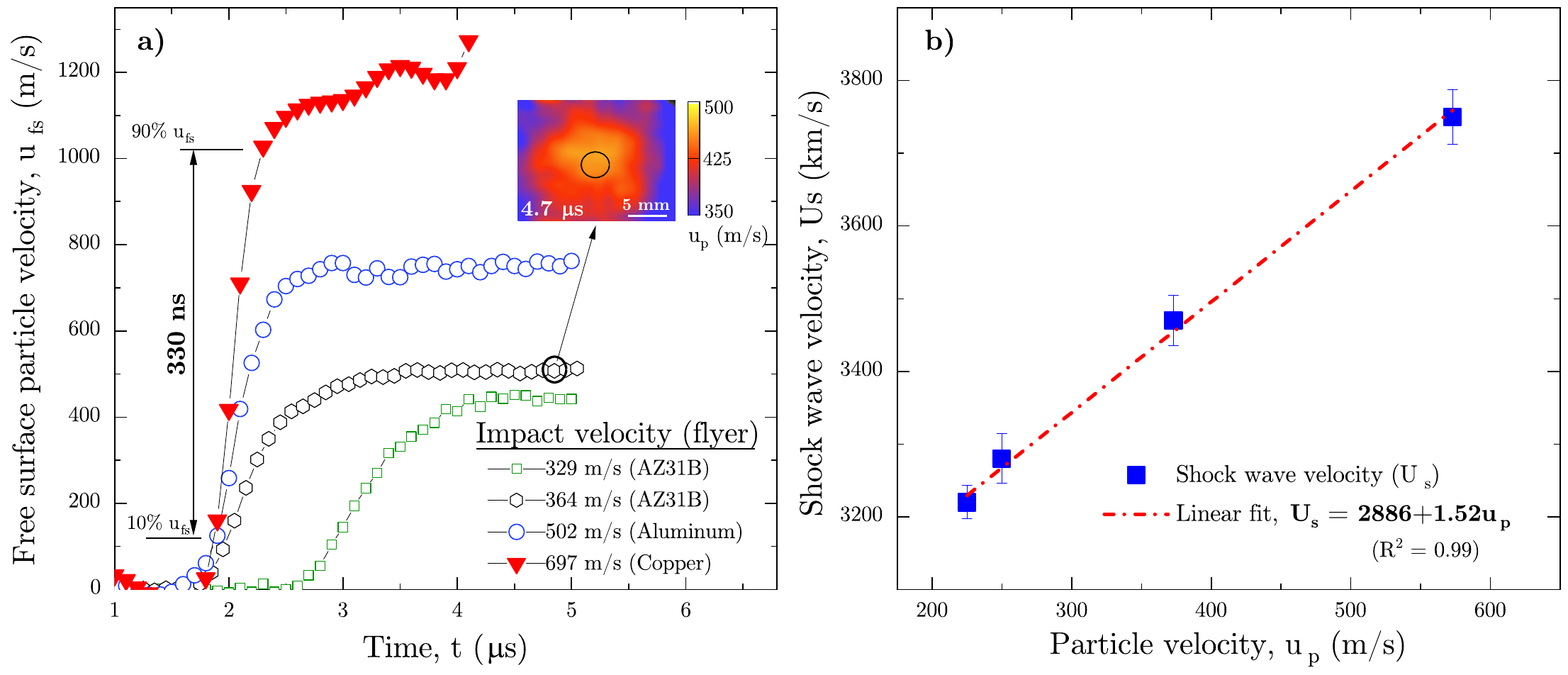}
	\caption{a) Free surface normal particle velocity at different impact velocities, and b) Shock wave velocity vs. particle velocity. }
	\label{fig:VelData}
\end{figure}

Normal plate impact experiments were conducted on the particulate composite samples at various impact velocities using different flyer materials to reach Hugoniot (Shock) stresses ranging from 1.0  to 3.0 GPa. The deformation images of the free surface acquired during the experiment were used to obtain free surface displacement fields and the full-field free surface particle velocity. In this study, the measurement was performed at the mesoscale since the experimental spatial resolution was close to the length scale of the glass particles in the composite. Therefore, to obtain the macroscopic free surface particle velocity, the velocity field was spatially filtered such that it contained at least five particles in the averaging window. An example of the spatially filtered full-field velocity contour for the experiment at 364 m/s (AZ31B impactor) is shown in the inset figure of Fig. \ref{fig:VelData}\hyperref[fig:VelData]{a}. 

The spatially filtered free surface particle velocity ($u_{fs}$=$2u_p$, where $u_p$ is the in-material particle velocity) at the center of the sample is plotted against time in Fig \ref{fig:VelData}\hyperref[fig:VelData]{a}. Here, time $t = 0$ $\mu$s corresponds to the contact time of the flyer with the target which is obtained from the copper shorting pins discussed in Section \ref{methods}. It is noted that the lowest velocity experiment, which is in the weak shock regime, has a typical ``lazy S'' shape particle velocity profile, which was also observed by Setchell et al. \cite{setchell2005shock}. The steady state free surface velocity for experiments Mg-GPC-1, Mg-GPC-2, Al-GPC, and Cu-GPC were approximately $450$, $500$, $746$ and $1150$ m/s, respectively and the corresponding particle velocities are half of the measured free surface normal particle velocity. For all the impact stresses, a clear rounding off region was observed after the initial particle velocity rise before stabilizing to the Hugoniot  state. These observations were also made by Rauls \cite{rauls2020structure}; however, the local deformation mechanisms for this overshoot remain unclear. As the impact velocity increases, the arrival time of the wave at the free surface decreases, implying higher shock velocity for faster impact velocities. At the same time, the steepness of the shock front drastically increases with the increasing impact velocity indicating a decrease in the shock width. While the equation of state (EOS) of each individual constituent in the composite is known, the EOS of the composite will be different and must be determined in order to establish the EOS of the composite. To obtain the EOS, the shock wave velocity was calculated by taking the thickness of the sample ($h$) and dividing it by the shock wave transit time in the sample. The shock wave transit time is the difference between the flyer contact time (corrected for tilt) and the shock wave arrival at the free surface of the sample. To consistently determine the arrival time in all the experiments, the time at which the free surface velocity attains 10\% of the Hugoniot state particle velocity was selected. This is because no clear distinction was observed between the elastic precursor and shock wave. Using the calculated shock wave velocities and Hugoniot state particle velocities, the $U_s-u_p$ relation was established and is plotted in Fig. \ref{fig:VelData}\hyperref[fig:VelData]{b}. It is noted that the EOS of the composite has a classical linear relationship ($U_s=C_0+Su_p$), which is seen in many engineering materials \cite{meyers_1994}. Once the $U_s-u_p$ relationship is established, it is essential to check whether the shock structure has fully developed, i.e. steady, using Bland's criterion \cite{bland1965shock}. The minimum thickness of the sample ($h$) required to develop a steady shock structure \cite{bland1965shock} is given by,
\begin{align}
    \Delta h=\frac{3 C_{0} U_{s} \Delta t}{8 S u_{p}} 
    \label{eq:Bland} 
\end{align}
where, $C_0$ and $S$ are the intercept and slope of the linear $U_s-u_p$ Hugoniot relation, and $\Delta t = $ rise time (time between 10 \% to 90 \%) of the particle velocity (see Fig. \ref{fig:VelData}\hyperref[fig:VelData]{a}). For the higher impact velocities, 502 m/s and 537 m/s, the minimum sample thickness required to generate a steady shock was 3.84 mm and 1.49 mm, respectively. While for the lowest impact velocity, the minimum thickness required to reach steady state shock was 9.3 mm, which is 0.3 mm larger than the thickness of the sample used in this study (9.03 mm). Therefore, to confirm the stability of the shock wave in the low-velocity impact case, an experiment was conducted with a 6.83 mm thick sample at a similar pressure using the same flyer material. The impact velocity measured in this experiment was 364 m/s, which was about 35 m/s higher than the previous experiment. Such differences in impact velocities are expected considering the nature of sabot acceleration in powder guns. However, the stresses in the sample should be in the vicinity of the prior experiment performed at 329 m/s since the difference in impact velocity is small. To check the shock structure, the rise time was compared between these two experiments. The measured rise time in the latter experiment was 900 ns which is very close to the rise time (1000 ns) measured for the experiment with the 9.03 mm thick sample. Note that the change in shock rise time was only 100 ns for a 2.2 mm sample thickness reduction even after considering slightly higher strength shock. Therefore, it is not expected that the shock structure and stability will be affected dramatically when the 9.03 mm thick sample is used in the experiment, in light of the fact that the data lies close to the linear fit of the $U_s-u_p$ relation.

The experiments were initially designed to reach stresses ranging from 1.0$-$3.0 GPa based on numerical simulations. However, using the particle velocity measurements from the experimental data and the EOS, the shock stresses were recalculated using,
\begin{align}
    \sigma=\rho_{0} U_{s} u_{p}
    \label{eq:stress}
\end{align}

\noindent where, $\rho_0$= initial density of the sample $(kg/m^3)$,  $U_s$= shock velocity $(m/s)$ is measured from the experiments, $u_p$ = one-half of the measured normal free surface velocity. All the quantities in Eq. (\ref{eq:stress}) are known, hence the shock stress for the lowest impact velocity was determined to be 1.0 GPa, and the highest stress achieved in this experiment was close to 3.1 GPa. The intermediate impact velocity experiment (impact velocity 502 m/s with aluminum flyer) yields normal stress close to 1.9 GPa. Shock strains ($\epsilon_{zz}$ = $u_p/U_s$) for the stresses 1.1 GPa, 1.9 GPa and 3.1 GPa are 0.069, 0.107 and 0.152, respectively, which are summarized in Table \ref{tab:Overview}.

\begin{table}[t]
 	\centering
 	\resizebox{0.98\textwidth}{!}{
 	\begin{threeparttable}
 		\setlength{\tabcolsep}{7.5pt}
 		\caption{Dimensions of flyer and sample, and summary of experimental results}
 		\centering
 		\footnotesize
 		
 		\begin{tabular}{cccccccccc} 
 			\hline \hline
 			\textbf{Experiment} & \begin{tabular}{@{}c@{}}\textbf{Flyer} \\ \textbf{{[}mm{]}}\end{tabular} & \begin{tabular}{@{}c@{}}\textbf{Target} \\ \textbf{{[}mm{]}}\end{tabular}& \begin{tabular}{@{}c@{}}\textbf{Impact} \\ \textbf{Velocity} \\ \textbf{{[}m/s{]}}\end{tabular} &  \begin{tabular}{@{}c@{}}\textbf{Normal} \\ \textbf{Stress} \\ \textbf{{[}GPa{]}}\end{tabular} & \begin{tabular}{@{}c@{}}\textbf{Rise} \\\textbf{Time}, $\Delta t$ \\ \textbf{{[}ns{]}}\end{tabular} & \begin{tabular}{@{}c@{}}\textbf{Steady} \\ \textbf{State} $u_{fs}$ \\ \textbf{{[}m/s{]}}\end{tabular} & \begin{tabular}{@{}c@{}}\textbf{Shock} \\ \textbf{Velocity}, $U_{s}$ \\ \textbf{{[}km/s{]}}\end{tabular} & \begin{tabular}{@{}c@{}}\textbf{Shock} \\ \textbf{Strain}, $\epsilon_{zz}$ \end{tabular} & \begin{tabular}{@{}c@{}}\textbf{Strain} \\ \textbf{Rate}, $\dot{\epsilon}$ \\ \textbf{{[}1/s{]}}\end{tabular}\\ \hline
 			Mg-GPC-1             & 4.890$\pm$0.003       & 9.03$\pm$0.007             & 329.3$\pm$0.3       & 1.1  & 1000 & 450 & 3.22&0.069 &85795        \\
 			Mg-GPC-2          & 4.986$\pm$0.003       & 6.23$\pm$0.020             & 364.6$\pm$0.2        & 1.2    & 900 & 500 &3.28 &0.076 &107166       \\
 			Al-GPC             & 9.806$\pm$0.005       & 6.89$\pm$0.01             & 499.2$\pm$0.5        & 1.9 & 600 & 746 &3.47 &0.107&253564  \\
 			Cu-GPC            & 4.969$\pm$0.002       & 7.34$\pm$0.004             & 697.8$\pm$0.2        & 3.1  & 330 & 1150 &3.75 &0.153&770560          \\ \hline \hline            
 		\end{tabular}
 		\label{tab:Overview}
 		\begin{tablenotes}
 			\footnotesize
 			\item **Flyer and target ring were 34 mm in diameter
 			\item **SLG bead diameter was 1 mm
 		\end{tablenotes}
 	\end{threeparttable}}
 \end{table}

\subsection{Rise time, shock thickness, and rate dependence}
\begin{figure}[ht]
	\centering
	\includegraphics[width=1\textwidth]{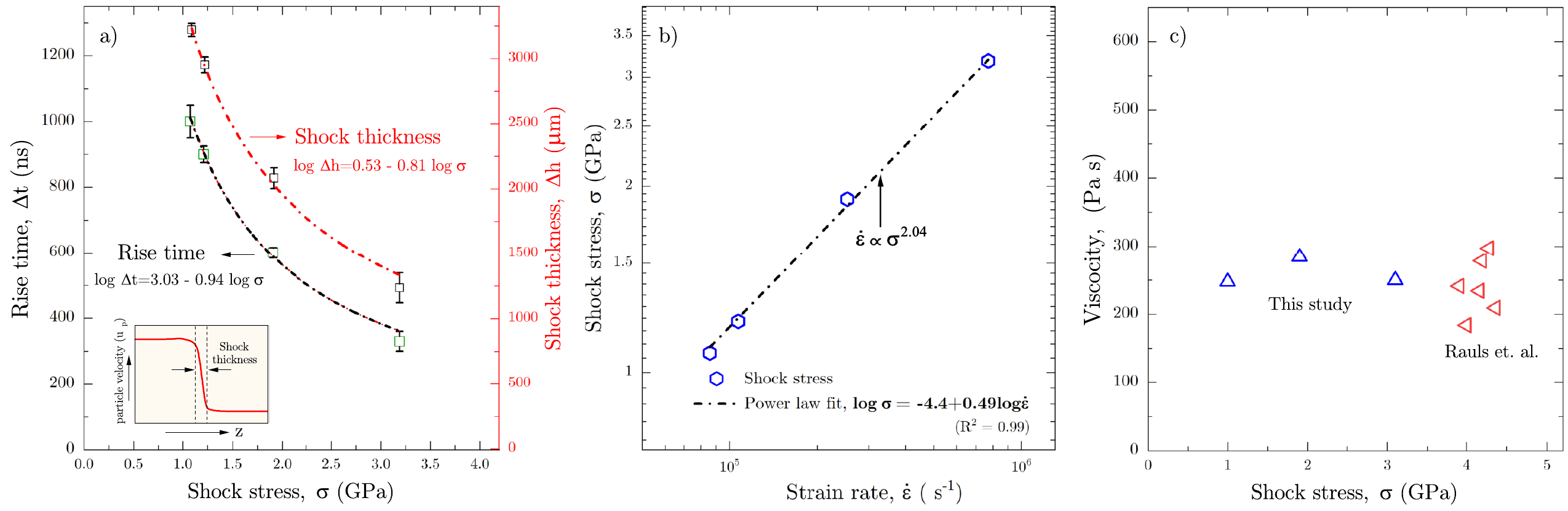}
	\caption{a) Rise time and shock thickness at varying shock stress, b) shock stress vs strain rate in log-log scale, linear fit with equation, $\log\sigma=-4.4+0.49 \log \dot{{\epsilon.}}$, c) shock viscosity at varying shock stresses. }
	\label{fig:ShockParams}
\end{figure}

In order to quantify shock thickness as a function of shock stress, the rise time was calculated (10/90 rise time, see Fig. \ref{fig:VelData}\hyperref[fig:VelData]{a}) for all four experiments. Figure \ref{fig:ShockParams}\hyperref[fig:ShockParams]{a} shows the variation of the rise time with shock stress. The shock rise time at $1.0$ GPa was $1,000$ ns. As the shock stress increases, the shock rise time decreases to 330 ns for a shock stress of 3.1 GPa. Conventionally, such a decrease in shock rise time with shock stress is associated with the decrease in shock viscosity at higher stresses. Based on the experimental results, the shock rise time variation with shock pressure appears to follow a power law, $\log \Delta t=3.03-0.94 \log \sigma$. Rauls et al.\cite{rauls2020structure} reported a rise time close to 225 $\pm$ $55$ ns for a similar composite (30\% volume fraction, $1000$ $\mu$m size glass spheres) at a shock stress of $4.1$ GPa. To compare their rise time measurement with the present study, the power law fit was extrapolated to 4.1 GPa. The rise time obtained was 280 ns at $4.1$ GPa, which is close to the rise time measured earlier \cite{rauls2020structure}. The small difference in the rise time measurement was attributed to the slight difference in volume fraction and uncertainties associated with the measurements. Using the the rise time $(\Delta t)$ and shock wave velocity $\left(U_{s}\right)$, the physical shock thickness was calculated using $\Delta h=U_{s}\Delta t$. Figure \ref{fig:ShockParams}\hyperref[fig:ShockParams]{a} shows the variation of the shock width with pressure. The physical width of the shock (schematically shown in the inset of Figure \ref{fig:ShockParams}\hyperref[fig:ShockParams]{a}) at $1.0$ GPa is around $3500$ $\mu$m (3.5 particles wide), which reduces to $1200$ $\mu$m (1.2 particles) at $3.1$ GPa. The width of the shock at the intermediate stress, $1.9$ GPa, is close to $2000$ $\mu$m (2 particles).

To establish the relationship between the strain rate $(\dot{\varepsilon})$ and stress jump $(\sigma)$, the strain rates were calculated using \cite{vogler2012scaling},

\begin{equation}
    \dot{\varepsilon}=\frac{\sigma}{\Delta t \rho_{0} U_s^{2}}
    \label{eq:StrRate}
\end{equation}

\noindent where all quantities in the equation were measured from the experiments. The shock stress as a function of strain rate is plotted in Fig. \ref{fig:ShockParams}\hyperref[fig:ShockParams]{b}, and it is apparent that the relationship between the strain rate and stress can be represented with a second power law relation, i.e $\dot{\varepsilon} \propto \sigma^{2.04}$. This is consistent with the plate impact experiments on similar particulate composites by Rauls et al. \cite{rauls2020structure}  and the study on layered composites by Zhuang et al. \cite{zhuang2003experimental}. It is interesting how both one-dimensional and three-dimensional composites lead to the same strain rate - stress scaling relation. To further evaluate the strain rate - shock stress relation, shock viscosity was also calculated using the relation $\eta=0.25 S \sigma \Delta t$ \cite{grady2010structured}. Based on this relation, for a constant shock viscosity, the rise time must vary inversely with the stress, i.e., $\Delta t \propto \sigma^{-1}$. Consequently, this implies the strain rate and shock stress must have a second power relation $\left(\dot{\varepsilon} \propto \sigma^{2}\right.$ ), see the equation for strain rate, Eq. (\ref{eq:StrRate}). Inspecting the rise time - shock stress power law relation from Fig. \ref{fig:ShockParams}\hyperref[fig:ShockParams]{b}, the power of stress was approximately $-0.94$, which is close to $-1$, indicating that the viscosity may be constant in the particulate composite considered in this study. The shock viscosity calculation for the wave profiles in this experiment yields nearly constant values $\sim 376,433,\textnormal{ and } 380$ Pa$\cdot$s for shock stresses, $1.1$, $1.9$, and $3.1$ GPa, respectively. Conventionally, the shock viscosity decreases with an increase in shock stress owing to nonlinear steepening mechanisms associated with the material. Therefore, to examine if such observation is consistent with the previous studies on this composite, the shock viscosity parameter $(\eta)$ was calculated using the data available in \cite{rauls2020structure}. Figure \ref{fig:ShockParams}\hyperref[fig:ShockParams]{c} shows the viscosity parameter for both the current study and previous study on glass filled composite and the viscosity does not seem to vary much with stresses. Therefore,  shock  front evolution may not be simply the action of material viscosity, but other mechanisms may also contribute to the shock wave structuring.

\subsection{Mesoscopic normal particle velocity}\label{MesoNormal}
\begin{figure}[ht]
	\centering
	\includegraphics[width=1\textwidth]{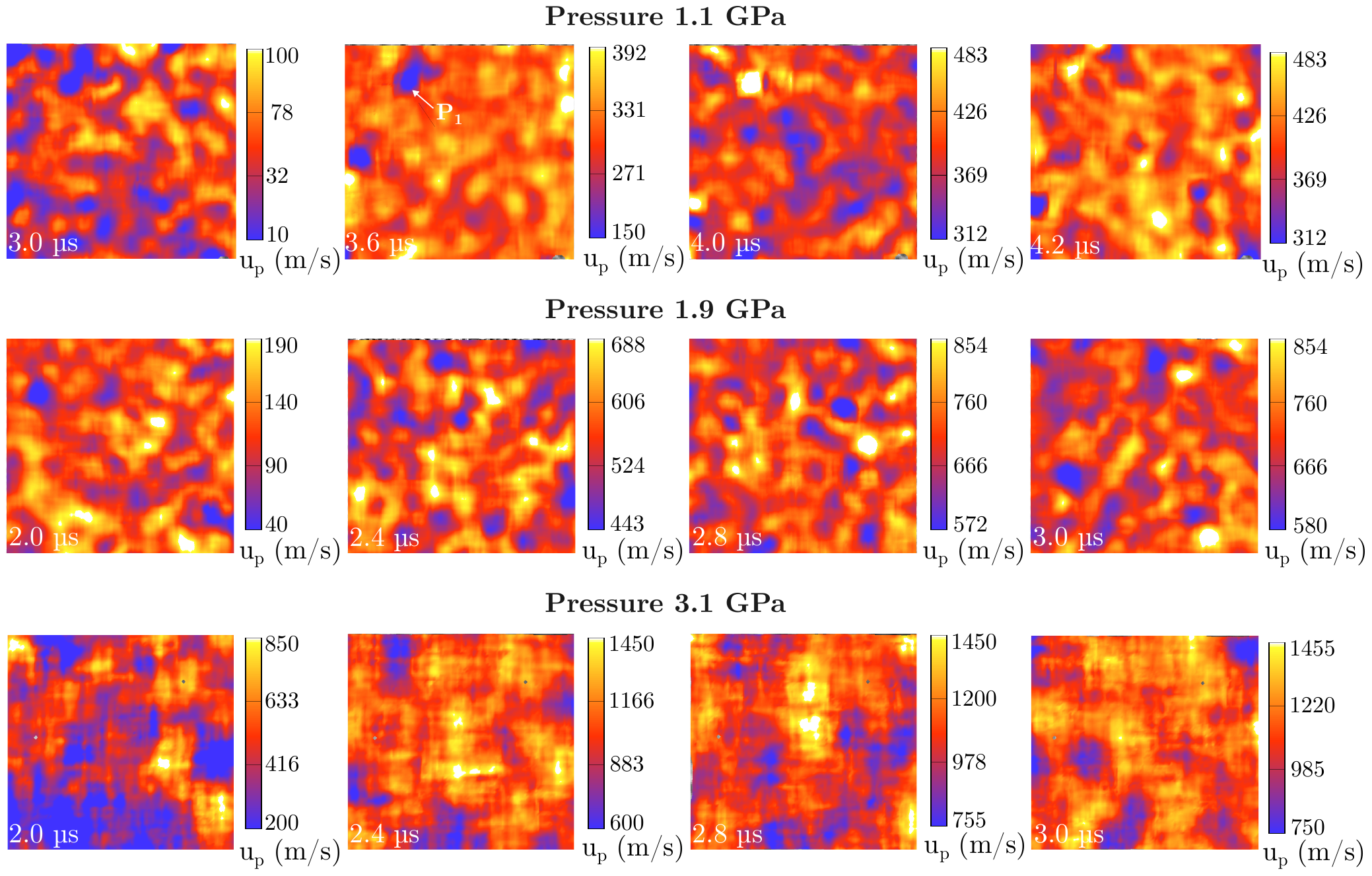}
	
	\caption{Mesoscale full-field normal particle velocity evolution at 1.1 GPa, 1.9 GPa and 3.1 GPa. The width of the window plot is $10$ mm $\times$ $10$ mm. }
	\label{fig:Meso}
\end{figure}

To understand the mesoscale free surface particle velocity and its effect in developing the shock structure, the particle velocity was computed from the displacement calculated using a small subset size $\left(1.2 \times 1.2 \mathrm{~mm}^{2}\right)$ and step size (68 $\mu$m) in Vic-3D. Spatial filtering of the data was applied using a non-local filter with a radius of 5 pixels, which resulted in a virtual gage length of $340$ $\mu$m. This small gage length allows to resolve the local deformation and particle velocity features in the composite material used in this study. Figure \ref{fig:Meso} shows the evolution of free surface normal particle velocity $\left(u_{fs} = 2u_p\right)$ at shock stresses varying from $1.1$ to $3.1$ GPa. Despite having a smooth rise and achieving a steady state shock front in the continuum scale, the mesoscale particle velocity $\left(u_{fs}\right)$ profile was highly heterogenous due to the mesoscale structure and the shock impedance mismatch between the constituents in the sample. For instance, in the $1.1$ GPa experiment, the average particle velocity profile was steady and close to $440 \mathrm{~m} / \mathrm{s}$ at the Hugoniot  state, see Fig. \ref{fig:VelData}\hyperref[fig:VelData]{a}. However as shown in Fig. \ref{fig:Meso}, the full-field mesoscale free surface velocity varies between $10$ m/s and $100$ m/s at $t=3.0$ $\mu$s (on the rise portion of the shock profile), while at the Hugoniot state ($t = 4.0$ $\mu$s), the velocity varies between 312 and $483$ m/s. At higher stresses and at $t = 3.0$ $\mu$s (at the Hugoniot state), the variation of the free surface velocity ranged from $580 - 854$ m/s for the $1.9$ GPa experiment and $750-1455$ m/s for $3.1$ GPa. Therefore, the shock front and the final Hugoniot state at the mesoscale appear to be spatially rough in contrast to the smooth wave profiles observed in the average normal particle velocity profile (Fig \ref{fig:VelData}). At higher pressures, the trend appears to be consistent; further, the heterogeneity does not change significantly with shock stresses. It was observed that some material points, which have low velocities initially, are shown to have an overshoot in velocity at a later time, indicating additional shock energy from the reflections at particle interfaces is added to these low-energy sites. For example, in the $1.1$ GPa experiment, point $\mathrm{P}_{1}$ had a velocity magnitude close to $150$ m/s at $t=3.6$ $\mu$s, while the majority of the other material points have velocities higher than $280$ m/s. However, at time $t=4.0$ $\mu$s, the velocity at point $\mathrm{P}_1$ peaks up to $483$ m/s, while most material points have velocities lower than $420$ m/s. Additionally, the velocity overshooting occurs near the steady Hugoniot state, indicating the reflected wave fronts may have caught up with the main shock front. These observations were made in all the experiments, indicating that the rough shock front tries to straighten itself using the additional energy obtained from the interface reflections. However, these free surface velocities are insufficient to deduce the type of reflections responsible for such catching-up behavior observed and this is discussed in later sections using numerical simulations.

\begin{figure}[ht]
	\centering
	\includegraphics[width=1\textwidth]{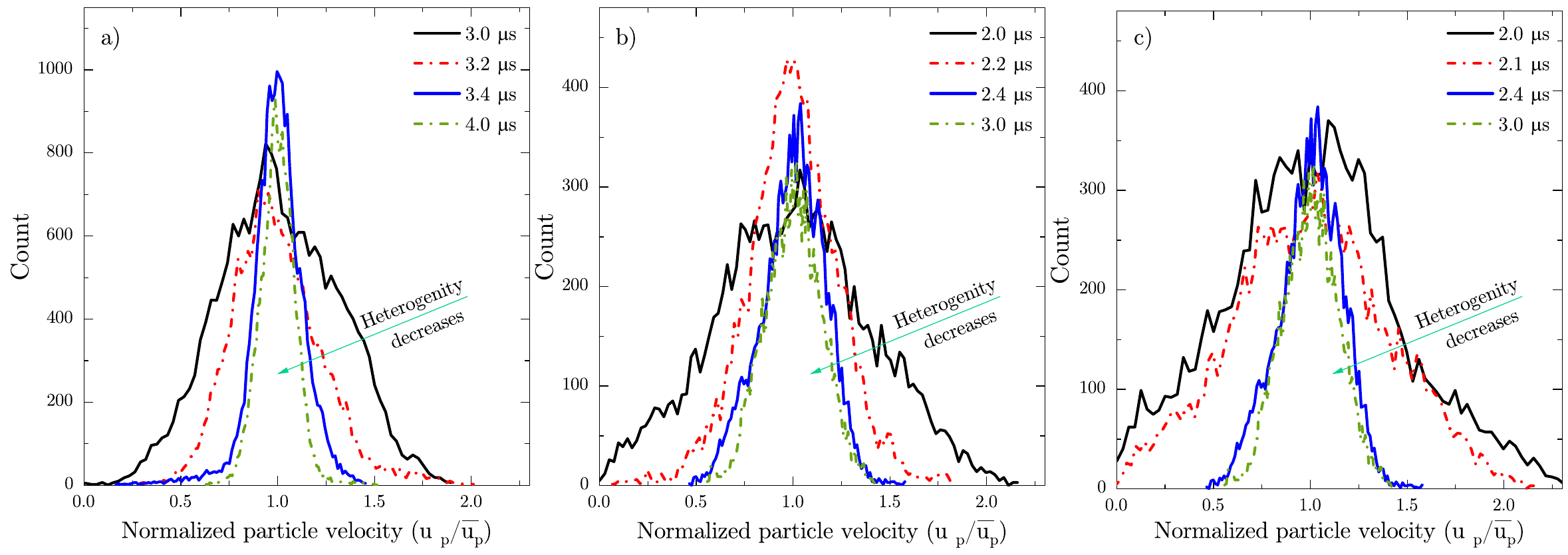}
	\caption{Frequency-normalized particle velocity histogram at different times for, a)  1.1 GPa, b) 1.9 GPa, c) 3.1 GPa. The particle velocity is normalized with respect to the mean particle velocity corresponding to the Hugoniot (steady) state.  }
	\label{fig:Heterogeneity}
\end{figure}

\begin{figure}[ht]
	\centering
	\includegraphics[width=1\textwidth]{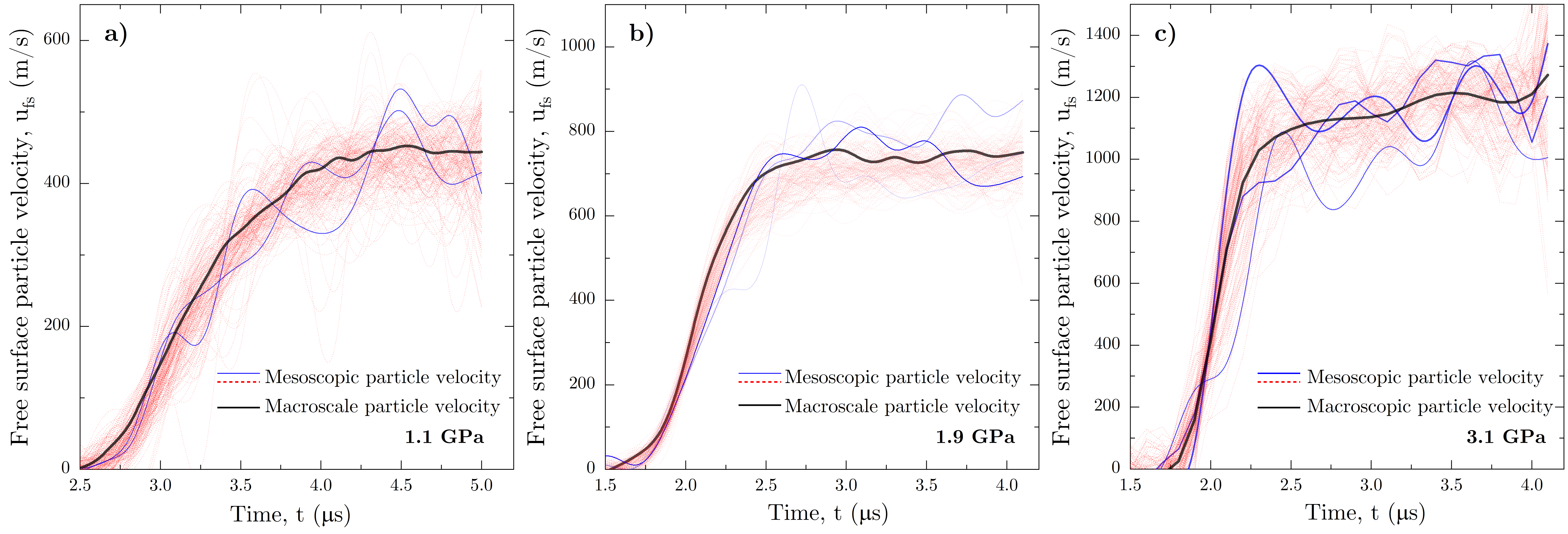}
	\caption{Free surface normal particle velocity for, a) 1.1 GPa, b) 1.9 GPa, c) 3.1 GPa. }
	\label{fig:ParticleVel}
\end{figure}

The normalized particle velocity is plotted against the frequency of observations at four time instances to quantify the spatial heterogeneity evolution in the material during shock propagation. The first two time instances were on the shock front (rise portion of the shock profile), while for the remaining two instances, one was during the rounding off region, and the other was at the peak stress state. Figure \ref{fig:Heterogeneity} shows the evolution of heterogeneity in particle velocity at three different stresses. The particle velocity (normalized with respect to the mean) distributions show a bell-shaped curve, and the width of the bell curve indicates the degree of heterogeneity present in the particle velocity. The particle velocity distributions are nearly symmetrically centered at one, indicating that the normal particle velocity below and above mean particle velocities are relatively the same. This indicates that the mechanism that decelerates or accelerates the shock wave being equally present, owing to the randomness in the microsctructure. For all the pressures, the heterogeneity is significantly higher during the shock rise than at the final peak stress state. For example, at 2.0 $\mu$s in Fig. \ref{fig:Heterogeneity}\hyperref[fig:Heterogeneity]{b}, the bell curve is wider compared to the bell curve at $t=2.2$ $\mu$s. Also, velocity heterogeneities are much smaller between the rounding-off region and the final Hugoniot state compared to the rising region. This change in heterogeneity suggests that within the shock front, the momentum diffusion process is highly heterogeneous due to the presence of mesoscale structure and high-rate deformation process. In contrast, behind the shock front, the velocity heterogeneity is smaller compared to the rising region, which is associated with the relaxation of the turbulent motion of the particles within the shock front after the shock passes. Contrary to the smooth behavior observed in the average profile (Fig. \ref{fig:VelData}), the Hugoniot state oscillates about a mean velocity, but with less heterogeneity than in the rising region. This observation is consistent in all pressures considered in this study, see Fig \ref{fig:ParticleVel}.

To study the observed spatial heterogeneity in the normal particle velocity field and its temporal evolution, the normal particle velocity vs. time for 100 points on the free surface of the sample is plotted in Fig. \ref{fig:ParticleVel}. At low shock stress, 1.1 GPa, the mesoscale shock appears to have multiple shock fronts (see blue curves) traveling at different velocities. In addition, the shock front shows characteristics of local stress release, possibly from the wave reflections at the interfaces. Therefore, spatial heterogeneity observed in the particle velocity in Fig. \ref{fig:Heterogeneity} is related to a train of shocks traveling at different velocities, which arrive at the free surface at different times. These weak disintegrated shock trains have been observed previously in shock wave propagation in a fluid with random particulates \cite{hesselink_1988}. The shock trains were attributed to the scattering of shock waves at the interfaces of the particles and fluid media. Also, several particle velocity profiles have shorter rise times compared to the average profile and arrive much earlier than the average profile, and vice versa. This steeper front indicates the presence of nonlinear steepening mechanisms primarily by the interference of the reflected waves from the interfaces. The arrival time of the shock at various points on the free surface varies anywhere between $0-300 \mathrm{~ns}$ in all the experiments. Some material points on the free surface show velocity overshoot, which is consistent with the previous experiments on similar materials \cite{rauls2020structure}. This overshoot is not a continuum characteristic of this material but rather related to its meso-structure. Therefore, multiple-point measurement of free surface velocities in a composite of this kind is required to describe the shock behavior accurately. To understand the shock structure evolution due to scattering, the Rayleigh elastic scattering equation is considered \cite{grady1998scattering},

\begin{equation}
    \alpha=\frac{4 \pi^3 v}{\lambda^4}\left\langle\left(\frac{\Delta E}{E}\right)^2\right\rangle_{\mathrm{ave}}
    \label{eq:scattering}
\end{equation}

\noindent here, $\alpha$ $=$ acoustic attenuation, $v=$ particle volume which scales as $a^3$ ($a=$ particle diameter), $\lambda$ $=$ wavelength of the plane wave, $\frac{\Delta E}{E}$ $=$ measure of property mismatch. Grady \cite{grady1998scattering} pointed out that the wavelength ($\lambda$) in the acoustic scattering attenuation relation carries a fourth power, which may explain the fourth power relation between strain rate and stress for the homogeneous materials. It is observed that in Eq. (\ref{eq:scattering}), there are two length scales, one is the particle diameter and the other is the characteristic wavelength. For shock compression experiments, wavelength refers to the shock thickness which is measured from the velocity data. Based on the attenuation coefficient relation (Eq. \ref{eq:scattering}), the scattering efficiency will be high if the shock width is higher than the particle diameter. Interestingly, this implies the widest shock front ($\lambda$ $=$ $3.2$ mm or $\sim 3$ SLG particles) observed in the low stress experiment will be most favorable for the scattering dominant shock structure evolution. However, the shock thickness presented here describes the width after the complete steepening of the front. Therefore, the shock front thickness could have been even higher at the beginning of the shock wave propagation which helps with wave scattering and shock structuring. In a scattering dominant field, the wave trains will be more pronounced since the shock energy is dispersed from the scattering of the waves such as the ones observed in the experiments. As the pressure increases, the scattering efficiency will decrease depending on the strength of shock. This may change the shock structuring mechanism from scattering dominant to viscosity driven. To answer if this is a possibility, higher pressure experiments are needed.

\subsection{Mesoscale transverse velocity }
\begin{figure}[ht]
	\centering
	\includegraphics[width=1\textwidth]{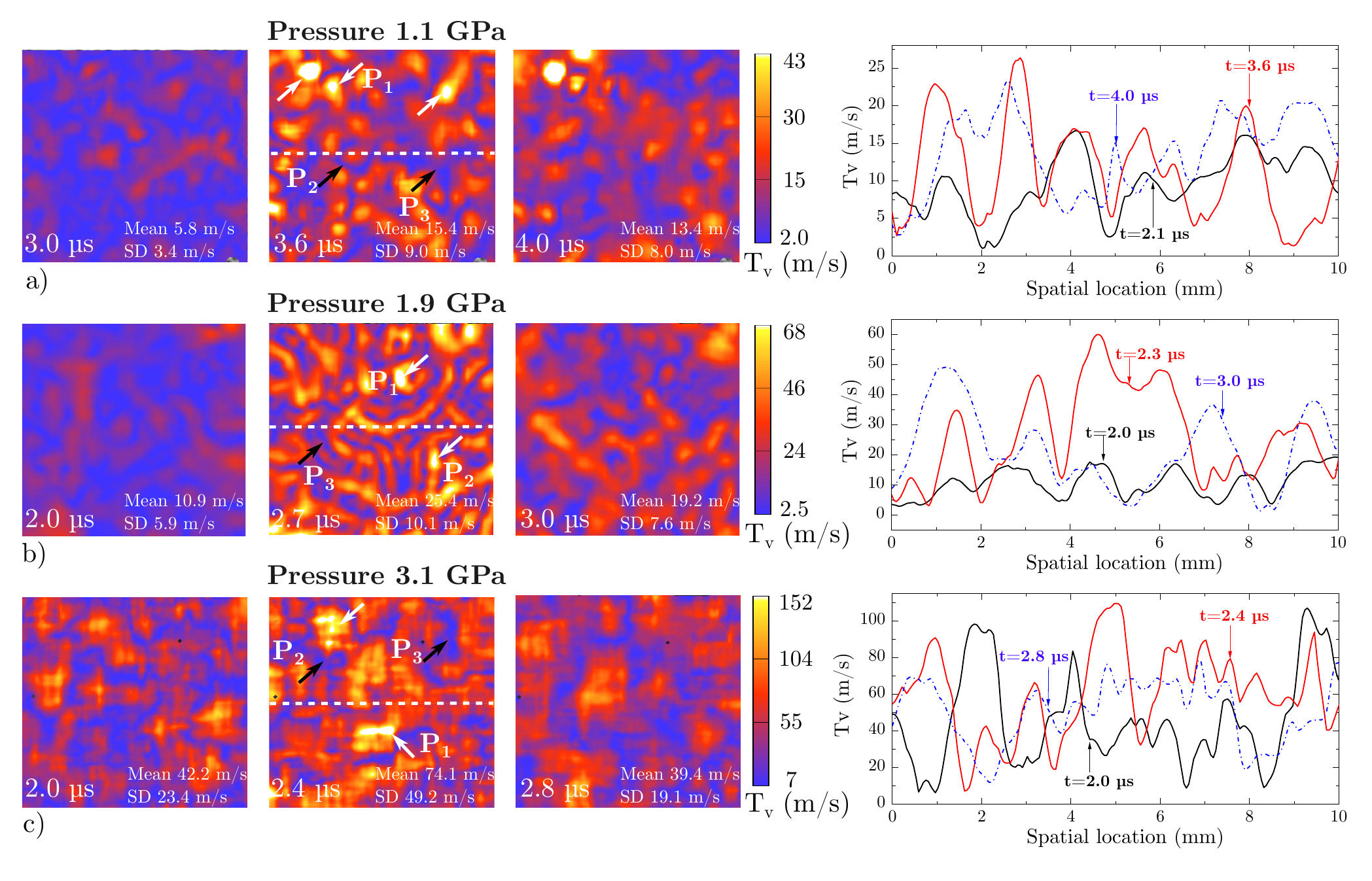}
	\caption{ Transverse particle velocity evolution and the spatial velocity heterogeneity along a line at the center of the sample for, a) 1.1 GPa, b) 1.9 GPa, c) 3.1 GPa. The window size of the contour plots is 10 mm $\times$ 10 mm.}
	\label{fig:Transverse}
\end{figure}

The spatial normal particle velocity heterogeneity observed in Section \ref{MesoNormal} indicates the presence of stress gradients in the material, which may cause momentum transport in all directions. To understand this momentum transport process in the lateral (transverse or in-plane) directions, the resultant transverse particle velocity is calculated using $T_{v}=\sqrt{V_{x}^{2}+V_{y}^{2}}$, where $V_{\mathrm{x}}$ and $V_{\mathrm{y}}$ are the in-plane particle velocities in $x$ and $y$ directions respectively. The contour plots and the spatial variation of the transverse particle velocity along a line at three time instances are plotted in Fig. \ref{fig:Transverse}. The first time instance corresponds to the shock rise region while the second time instance describes the rounding-off region of the shock profile, and the final time instance is at the Hugoniot state. The full-field particle velocity contours demonstrate a heterogeneous transverse particle velocity evolution, indicating a highly heterogeneous lateral momentum diffusion process at the mesoscale. The spatial velocity variations along a line at the center exhibit ripples of different amplitudes. During the initial shock rise (within the shock front), the average amplitudes of the ripples are low, owing to low normal particle velocity (see spatial velocity profile). As the normal particle velocity increases, the amplitude of the velocity of the ripples increases up until the rounding-off region, and then it stays nearly steady. For instance, in the $1.1$ GPa experiment, the average amplitude of the ripples at $t=2.1$ $\mu$s (during shock rise) is below $3.4$ m/s, but increases to $9.0$ m/s at $t=2.7$  $\mu$s and remains at around $8$ m/s afterwards. Similar behavior can be observed for the $1.9$ GPa and $3.1$ GPa experiments. 

\begin{figure}[ht]
	\centering
	\includegraphics[width=1\textwidth]{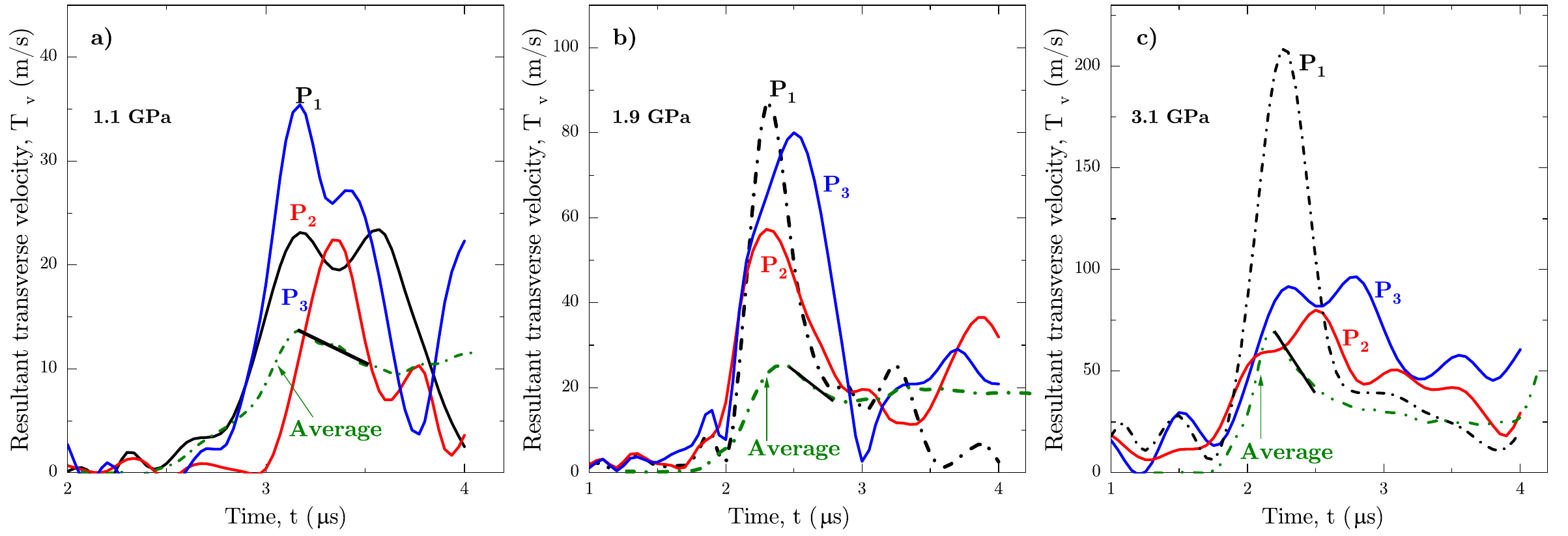}
	\caption{Resultant transverse velocity evolution for, a) 1.1 GPa, b) 1.9 GPa, c) 3.1 GPa }
	\label{fig:TvEvo}
\end{figure}

The transverse particle velocity evolution at three materials points $\left(\mathrm{P}_{1}, \mathrm{P}_{2}\right.$ and $\left.\mathrm{P}_{3}\right)$ and the average transverse particle velocity evolution for all three experiments are plotted in Fig. \ref{fig:TvEvo}. The transverse particle velocity rises rapidly during the shock rise and after attaining a peak velocity in the rounding-off region, it gradually decreases to nearly a constant magnitude. Additionally, the peak transverse particle velocity at different points occurs at different times, which is related to the varying arrival time of the normal shock observed in Section \ref{MesoNormal}. The rapid rise in particle velocity in the transverse direction leads to high local gradients in shear stresses. Therefore, the strength properties of the matrix material should have an important role in structuring the shock wave. Also, the highest rate of change of momentum diffusion in the lateral direction occurs during the rise time of the shock (within the shock front) and is apparent from the large slope during the rise in the transverse particle velocity evolution. This redistribution of energy at the mesoscale can cause the formation of a wider shock front at a given stress, and leads to the increased viscosity resulting in a smaller exponent for the relation between strain rate and stress as observed in this study. After the shock rise, the rate of momentum diffusion decreases in the rounding-off region, followed by a gradual transition to a steady state. For instance, in the experiment at $1.1$ GPa shock stress, the average transverse velocity rises to $15.4$ m/s up to the rounding-off region (see Fig. \ref{fig:TvEvo}\hyperref[fig:TvEvo]{a}) and then relaxes to $10$ m/s. While in the $1.9$ GPa experiment, the average particle velocity rises to $28$ m/s and relaxes to $22.2$ m/s as seen in Fig. \ref{fig:TvEvo}\hyperref[fig:TvEvo]{b}. For the highest velocity experiment, the relaxation occurs from $84$ m/s to $39$ m/s. The peak transverse momentum diffusion is about $3.4\%$ of the normal particle velocity for the $1.1$ GPa experiment, while it increases to $4.1\%$ and $7.3\%$ for the $1.9$ and $3.1$ GPa experiments respectively. Conventionally in homogeneous materials \cite{grady1981strain} as the shock stress increases, the shock thickness decreases rapidly, which leads to the fourth power relation between strain rate and stress. Similarly, in the particulate composite, as the impact stress increases the shock thickness decreases, but at a much weaker rate, thus leading to a weaker second power strain rate-stress power law relation. Interestingly, the lateral momentum diffusion increases with increase in impact stress, which allows the impact energy in the longitudinal direction to redistribute in the lateral direction. This redistribution of the shock energy causes an increase in the shock width, which could be one of the mechanisms for the observed weaker, second power relation. The slope of the transverse velocity relaxation (see the black line on the average velocity profile of Fig. \ref{fig:TvEvo}) appears to increase with increase in stress, attributing to quicker attainment of steady-state particle velocity in high-stress experiments. For example, in the highest stress experiment, the transverse velocity rises to $72$ m/s, but quickly relaxes to $39$ m/s in about $400$ ns. The slope of the relaxation curve (deceleration) is $7.1 \times 10^{8} \mathrm{~m} / \mathrm{s}^{2}$. While in the $1.1$ GPa and $1.9$ GPa experiments, the relaxation slope is much lower, close to $7.0 \times 10^{6} \mathrm{~m} / \mathrm{s}^{2}$ and $1.2 \times 10^{7} \mathrm{~m} / \mathrm{s}^{2}$, respectively. It is noted that such slow redistribution of energy during relaxation can lead to a slower transition to the final Hugoniot state causing the normal particle velocity rounding-off observed in this study. Consequently, a sharper rounding-off was observed in the $3.1$ GPa experiment compared to the $1.1$ GPa and $1.9$ GPa experiments.

\subsection{Numerical Simulation: normal particle velocity}
\begin{figure}[ht]
	\centering
	\includegraphics[width=0.8\textwidth]{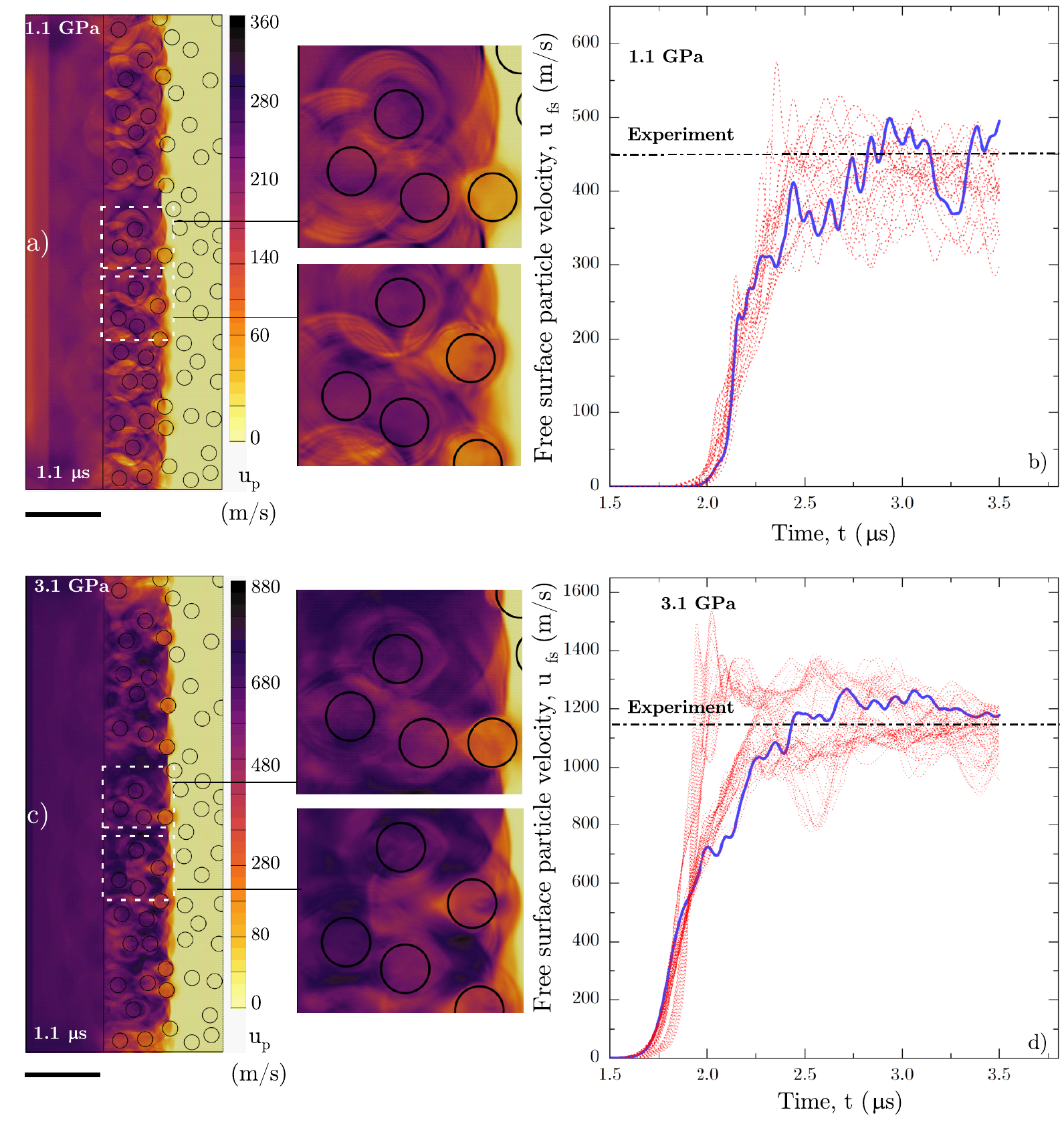}
	\caption{a) Overlaid contour plot of normal particle velocity on the microstructure for the 1.1 GPa simulation, and b) the evolution of free surface normal particle velocity. Similarly, c) overlaid contour plot of normal particle velocity in 3.1 GPa simulation, and d) its evolution of the free surface normal particle velocity. Scale bar is $5$ $mm$.}
	\label{fig:Simulation}
\end{figure}

Numerical simulations of two-dimensional particulate composites were conducted at two stresses, $1.1$ and 3.1 GPa, to gain insights into the mechanisms responsible for the observed heterogeneous shock structure. A random distribution of particles of 23\% volume fraction, similar to the experiments, was generated using an algorithm written in Python. Both simulations were conducted using a bonded interface between the inclusions and the PMMA matrix, and damage models for the particulates or the matrix were not incorporated. Hence, the delamination and relative sliding of the interfaces and failure of the matrix or the particles were not appropriately modeled. For both simulations, the same microstructure and dimensions were used to compare the results. Figure \ref{fig:Simulation}\hyperref[fig:Simulation]{a} and \ref{fig:Simulation}\hyperref[fig:Simulation]{b} depict the normal velocity contour plot overlaid on the microstructure in addition to the zoomed view of the contours. Also, the free surface particle velocity evolution for 100 material points on the free surface is plotted in Fig. \ref{fig:Simulation}\hyperref[fig:Simulation]{c} and \ref{fig:Simulation}\hyperref[fig:Simulation]{d}. The contour plots show highly spatially heterogeneous velocity fields with specific local particle velocity features. In the low-stress experiment at $1.1$ GPa, when the wave arrives at the particle, distinct trough and crests were observed as a sign of the scattering of the waves, shown by the white arrows in the magnified region in Fig. \ref{fig:Simulation}\hyperref[fig:Simulation]{a}. However, these troughs and crests were not observed in the experiment at 3.1 GPa, which supports the hypothesis that lower scattering occurs at high stresses due to smaller shock width (wavelength). The normal free surface particle velocity plots show a highly heterogenous particle velocity evolution with a velocity varying from 300 to $550$ m/s after the initial rise for the $1.1$ GPa impact, qualitatively very similar to the experiments. Similarly, in the $3.1$ GPa simulation, the normal particle velocity varies between $800$ to $1500$ m/s matching the oscillations observed in the experiments. The shock velocity calculated for these simulations were approximately $3.3$ and $4.1$ km/s for the $1.1$ and $3.1$ GPa, respectively. The peak Hugoniot state closely matched with the experiment, however, the rise time in the 1.1 GPa simulation appeared to be shorter, possibly due to the result of several local deformation mechanisms not captured or inactive due to low confinement expected under low stresses. The ripples observed in the simulations along with the weak shock trains and the varying shock arrival times are qualitatively similar to the mesoscale experimental measurements in this study. Most of the characteristics of the shock wave in the particulate composites were captured in the simulation with the simple phenomenological models used in this study.

\begin{figure}[ht]
	\centering
	\includegraphics[width=1\textwidth]{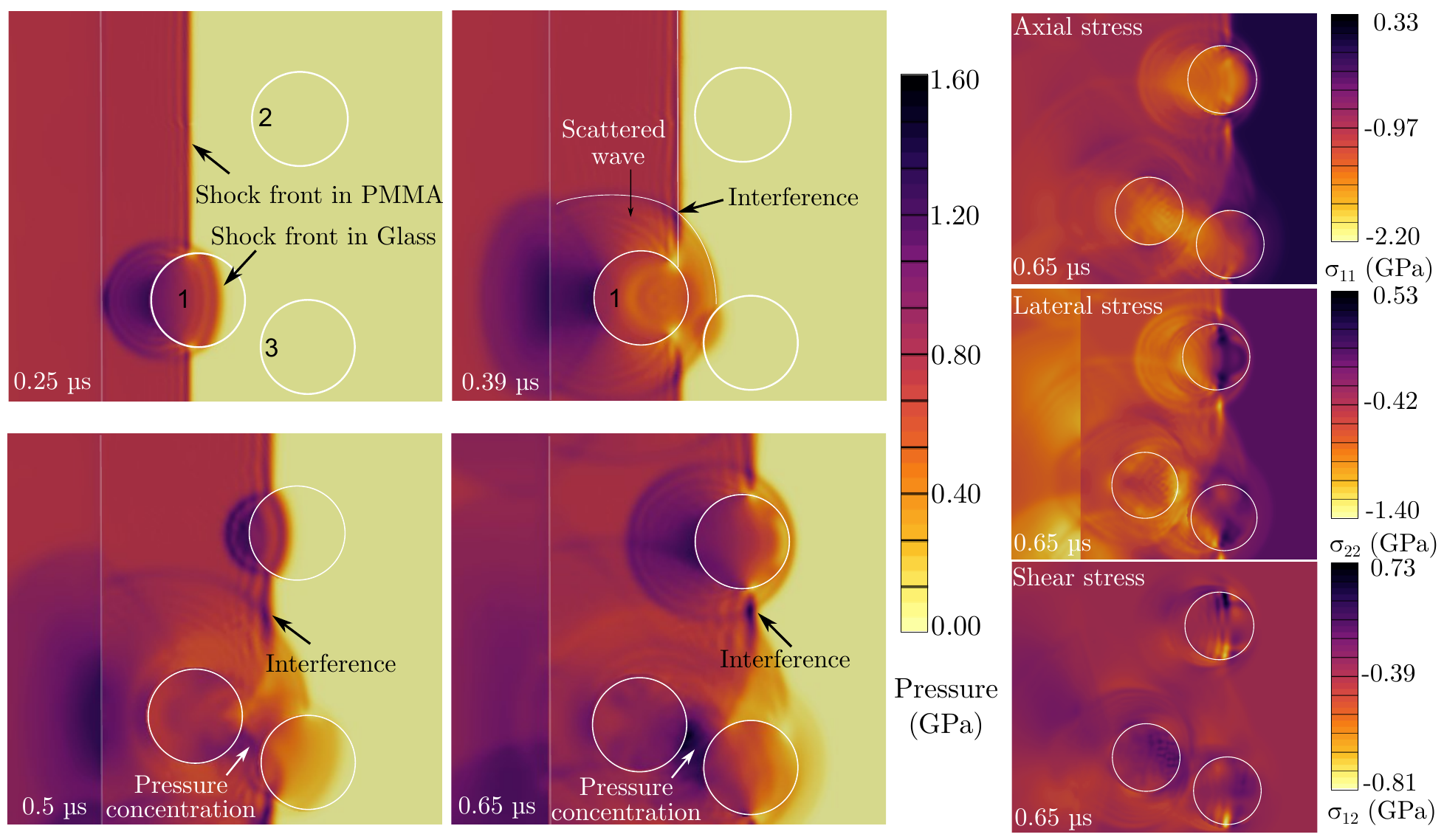}
	\caption{Zoomed view of the local pressure evolution contour plots at 1.1 GPa impact stress. Additionally, the axial, lateral, and shear stress contours are shown at  $t = 0.65$ $\mu$s. }
	\label{fig:SimMech}
\end{figure}

To investigate the mesoscale mechanisms for the shock front roughness and the oscillations in the shock profile, the local pressure contours in $1.1$ GPa simulation were examined in detail. The zoomed view of the pressure contours in a region close to the impact end is shown in Fig. \ref{fig:SimMech}. At time, $t=0.25$ $\mu \mathrm{s}$, the shock front in the PMMA lags behind the shock in the glass inclusion-1 due to the higher shock speed in the glass. Therefore, the arrival time of the shock in the path involving a greater number of particles will have a shorter shock transition time and vice versa, which explains the wide range of shock arrival times observed in the experiments. Also, the reflected wave from the interface between the glass bead-1 and the PMMA causes a rise in local pressure behind the particles due to the higher impedance of the beads. While reflections between the particles, for instance between 1 and 3 glass beads at $t=0.65$ $\mu$s, cause the local pressure to be significantly localized. The pressure near this interface is close to $1.4$ GPa, whereas in the PMMA matrix away from the interface, the stress is $0.4$ GPa. The local normal and lateral stresses in the region between the beads are compressive with magnitudes between $2.2$ and $1.40$ GPa. It is noted that such interfaces have high shear stresses, close to $0.73$ GPa, therefore these types of reflections have a role in viscous flow-driven shock structure evolution. Interestingly, the scattered wave from the particle propagates in the diagonal direction, see $t=0.39$ $\mu$s, which interferes with the forerunning plane shock in the PMMA matrix and increases the local pressure/stress. This interference of scattered waves with the plane shock wave produces a similar effect of shock focusing. The pressure at the interference location is $1.6$ GPa, the highest pressure observed at the time instance. Several locations are observed to have positive stress indicating relaxation due to the components of the stress scattered in opposite directions of the shock propagation and possible destructive interference. These oblique scattered waves carry high local shear stresses as seen in the shear stress plot at, $t=0.65$ $\mu$s, which will have a dominant role in the shock structure evolution. Scattering, shock focusing due to interference, and shock reflections at the interfaces can magnify the local particle velocity, while stress relaxation and the destructive interference can locally relax the stresses in the material. This shock strengthening and weakening are the reasons for the oscillating velocity profiles even after reaching the Hugoniot state. Also, the high spatial gradient observed in the local stresses allows the material to laterally deform during the shock propagation, which realigns the particles and diffuses the momentum in the lateral directions causing the rounding off the shock profile as seen in the normal velocity measurements (Fig. \ref{fig:Simulation}).
\section{Summary and Conclusions}

Normal plate impact experiments, coupled with high speed imaging and stereo digital image correlation (DIC), were conducted on model particulate composites to characterize their heterogeneous mesoscale response for the first time. This study explored the fundamental mesoscale mechanisms for the observed power law scaling between stress and strain rate. The key results and observations are summarized below:
\begin{enumerate}
    \item A spatial filtering of the free surface velocity was conducted to extract the average homogeneous response at the center of the composite material. Similar to most engineering materials, a linear $U_s-u_p$ relation was obtained, and a decrease in the shock rise time was observed with increasing impact stress. While for metals, this decrease is associated with the decrease in viscosity at higher stresses, an alternative mechanism seems to play a role for particulate composites. Interestingly, an overshoot of the average velocity was observed at higher stresses which can be explained through the observed mesoscopic response. 
    \item The scaling between the rise time and shock stress was found to follow the relation, $\Delta t\propto \sigma^{-0.94}$. This scaling implies an approximately constant observed shock viscosity however, the physical implications are subject to future investigation. Similar to Rauls et al. \cite{rauls2020structure}, and Zhuang et al. \cite{zhuang2003experimental}, the strain rate scales as approximately the square of normal stress ($\dot{\varepsilon} \propto \sigma^{2.04}$), which indicates a significant dissipative process within the shock front.
    \item The mesoscopic normal free surface velocity was determined across the entire rear surface of the shocked composite using DIC analysis. Significant spatial and temporal heterogeneity for the normal (out-of-plane) velocity was observed due to wave scattering from the difference in the impedance of the PMMA and SLG beads. The highest spatial heterogeneity was observed during the rise portion of the shock, indicating a heterogeneous momentum diffusion process occurring due to the high rate deformation process. This spatially and temporally oscillating velocity field is associated with shock trains generated due to interface reflections, scattering from particles, and interference with other reflected shocks, which travel at varying velocities and coalesce with the main shock front. This describes the multiple shocks observed during the rise for the 1.1 GPa experiment indicating higher scatter at lower stresses, and also explains the overshoot observed at higher shock stresses as the roughness of the wave increases at higher impact stress. Unlike viscosity driving the shock structure, like for homogeneous materials, the dominant mechanism for particulate composites seems to be wave scattering from these inclusions. 
    \item While normal impact does not generate any transverse velocity, a highly heterogeneous in-plane transverse velocity was also observed in the experiments as a result of the heterogenous deformation process at the mesoscale. The transverse motion indicates a heterogeneous lateral momentum diffusion which was the strongest during the rise of the shock wave and is proportional to the impact stress. The redistribution of energy along the transverse direction is critical for widening the shock front (i.e., shock width) and potentially explains the weak relation between strain rate and normal stress. Additionally, it is possible that transverse diffusion is strongly related to the strength of the PMMA binder as the rapid change in transverse velocity is directly proportional to the shear stress in the composite.
    \item To better explain the scattering mechanisms due to the arrangement of the beads, a series of 2D finite element simulations were conducted on the PMMA matrix with randomly distributed soda lime glass beads. Constructive and destructive interference of scattered waves from the soda lime glass particles were observed, explaining the oscillating free surface velocity at the Hugoniot state. Ultimately, the constructive interference results in mechanisms such as shock focusing, which magnifies the local particle velocities, causes high-stress concentrations, and strengthens the propagating shock. Consequently, the scattered waves' destructive interference relaxes stresses and forms regions of local tension in the material. These dissipative mechanisms define the shock structure and the weak relation observed between stress and strain rate.

\end{enumerate}

\section*{Acknowledgements}
The research reported here was supported by the DOE/NNSA (Award No. DE-NA0003957), which is gratefully acknowledged. The authors acknowledge the Army Research Laboratory (Cooperative Agreement Number W911NF-12-2-0022) for the acquisition of the high-speed cameras.

\bibliographystyle{elsarticle-num}
\bibliography{References.bib}

\end{document}